\documentclass[amsmath,aps,prl,preprint,showpacs,byrevtex]{revtex4}
\usepackage{graphicx}

\begin{document}

\title{The origin of fat tailed distributions in financial time series}

\author{G. M. Viswanathan$^1$, U. L. Fulco$^{1,2}$, M. L. Lyra$^1$ and
M. Serva$^{1,3}$}\vspace{2cm}

\affiliation{$^1$ \mbox{Departamento de F\'{\i}sica, Universidade Federal de
 Alagoas,}\\  Macei\'{o}--AL, 57072-970, Brazil}

\affiliation{$^3$  Departamento de F\'{\i}sica, Universidade Federal de
 Piau\'\i, Teresina--PI,  64049-550 , Brazil}

\affiliation{$^3$  Dipartimento di Matematica and I.N.F.M.
Universit\`a dell'Aquila, I-67100 L'Aquila, Italy}

\revised{06 November 2002}

\vspace{2cm}

\begin{abstract}
A classic problem in physics is the origin of fat tailed distributions
generated by complex systems.  We study the distributions of stock
returns measured over different time lags $\tau.$ We find that
destroying all correlations without changing the $\tau = 1$~d
distribution, by shuffling the order of the daily returns, causes the
fat tails almost to vanish for $\tau>1$~d.  We argue that the fat
tails are caused by known long-range volatility correlations.  Indeed,
destroying only sign correlations, by shuffling the order of only the
signs (but not the absolute values) of the daily returns, allows the
fat tails to persist for $\tau >1$~d.  
\end{abstract}

\maketitle

Over the last few decades, remarkable progress has been made in
quantitatively describing non-Gaussian phenomena, including those
observed in economic~\cite{econophysicsbook} and
social~\cite{montroll} systems, that are typically characterized by
the presence of fat tailed L\'evy distributions~\cite{shlesinger}.
The behavior of financial markets has
recently~\cite{econophysicsbook,prl1,prl2,plamen1,plamen2,lrc1,lrc-old0,gopi,lrc2,mantegna2,mantegna}
become a focus of interest to physicists as well as an area of active
research because of its rich and complex
dynamics~\cite{mandelbrot,kadanoff,mantegna3,Li,takayasu,bak,bouchaud,ghashghaie,levy,galluccio,bak2,vandewall,Liu,cizeau,volatility-stanley,ingersoll}.
One open question relates to the probability distribution underlying
returns on stock markets.  It is well known that the century-old
Gaussian model~\cite{bacchelier} underestimates the probability of
large events.  Indeed, the distribution of stock returns is fat
tailed~\cite{econophysicsbook,mantegna,mandelbrot}.  On the one hand,
the fat tails could be due to an underlying L\'evy distribution.
According to the generalized central limit theorem, the sum of $\tau$
independent (i.e., uncorrelated) L\'evy distributed random variables
is also L\'evy distributed, such that the persistence of fat tails for
large $\tau$ is due solely to the L\'evy nature of the original
($\tau=1$) distribution.  Furthermore, if the $\tau=1$ distribution is
L\'evy with exponential truncation of the tails, then we still expect
a certain degree of persistence of the fat tails~\cite{tlf}. On the
other hand, fat tails can also persist~\cite{gopi} for $\tau>1$ due to
long-range correlations in a ``hidden variable''~\cite{lrc1} such as
volatility (i.e., locally measured standard deviation)~\cite{plamen1}.
Moreover, such long-range correlations have been found to produce fat
tails~\cite{volatility-stanley}.  Although stock returns lack
long-range power law correlations, yet the absolute values of the
returns are known to be long-range
correlated~\cite{lrc1,lrc-old0,lrc2,lrc-old1,lrc-old2,lrc-old3,lrc-old4}.
The absolute returns are power law correlated with non-unique scaling
exponents~\cite{lrc1,lrc2}.  Here we test the hypothesis (see
ref.~\cite{gopi}) that long-range volatility correlations are the
origin of the fat tails.

Below we will show that for stock market returns the observed
persistence of fat tails for large $\tau$ cannot be explained without
long-range correlations in the volatility.  Shuffling the daily
returns has the effect of destroying all correlations while
maintaining unchanged the $\tau=1$~d distribution.  For shuffled data,
we will show that the distribution is fat tailed for $\tau=1$~d but
not for $\tau>1$~d.  We interpret this finding as evidence that
volatility correlations rather than the $\tau=1$~d L\'evy-like
distribution are responsible for the existence of fat tails for large
$\tau$.  Indeed, we will also show that shuffling only the signs of
the returns allows the fat tails to persist for $\tau>1$~d and the
distribution does not converge to a Gaussian.  These findings
conclusively prove that known long-range volatility correlations
(rather than known short-range~\cite{mantegna2} sign correlations) are
responsible for fat tails for any $\tau>1$~d.  We will also show that,
remarkably, the short-range (1--2~d) sign correlations also play an
important role in the distribution properties of small price changes,
and we will propose an explanation for why a Gaussian fits the data so
well for small, but not large, returns.

Our dataset consists of the base 10 logarithms of daily returns
obtained from 59 stock market indices (obtained from yahoo.com: AEX,
AORD, ATG, ATX, BFX, BSESN, BVL30, BVSP, CCSI, DJA, DJI, DJT, DJU,
DOT, FCHI, FTSE, HEX, HSI, IBC, IGRA, IIX, IPSA, IXIC, JKSE, KFX,
KLSE, KS11, KSE, MERV, MID, MTMS, MXX, N225, NDX, NTOT, NYA, NZ40,
OEX, PSE, PSI, PX50, RUA, RUI, RUT, SAX, SETI, SML, SMSI, SOOX, SPC,
SSEC, SSMI, STI, TA100, TSE, TWII, VLIC, XMI, XU100).  The returns
$r(t)$ are defined in terms of the prices $P(t)$ by
\begin{equation}
r(t)\equiv 
\log_{10}\frac{P(t)}{P(t-1)}\;\;.
\end{equation}
We normalize the returns to unit variance for each market index
separately.  To be able to compare returns measured over differing
time scales, we also define a rescaled return $r_\tau$ by
\begin{equation}
r_\tau(t)\equiv \frac{1}{\sqrt{\tau}}
\sum_{t'=t-\tau+1}^{t} r(t')\;\; ,
\end{equation}
with $\tau$ measured in days and $r_1=r.$ The daily and rescaled
returns play roles similar to those played by ``bare'' and ``dressed''
quantities in field theory. Note that for uncorrelated (independent)
and unitary Gaussian distributed returns, their variances will be
identical due to the central limit theorem:
$\sigma(r_1)=\sigma(r_\tau)=1$~d. Similarly, if $r_1(t)$ are L\'evy
distributed, then $r_\tau(t)$ will also be L\'evy distributed.

\begin{figure}
~
\bigskip
~ 
\centerline{\includegraphics[width=12cm,angle=-90]{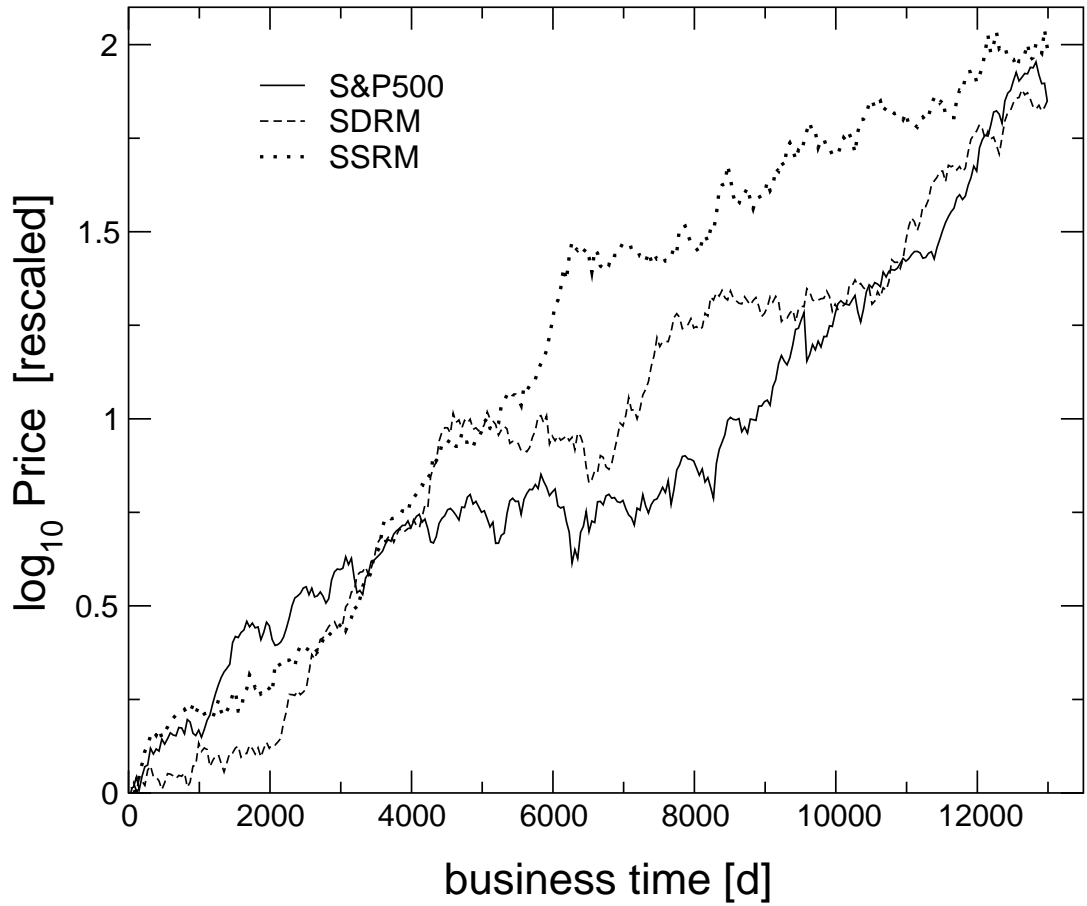}}
\medskip\medskip
\medskip

\caption{S\&P~500 index, shown on a base 10 logarithmic scale, offset
to zero. Also shown are SDRM which has completely uncorrelated returns
but an identical $\tau=1$~d distribution, and SSRM, which has an
identical probability distribution of the absolute daily returns, but
lacks sign correlations.  }
\label{sp}
\end{figure}

\begin{figure}
\centerline{\includegraphics[width=12cm,angle=-90]{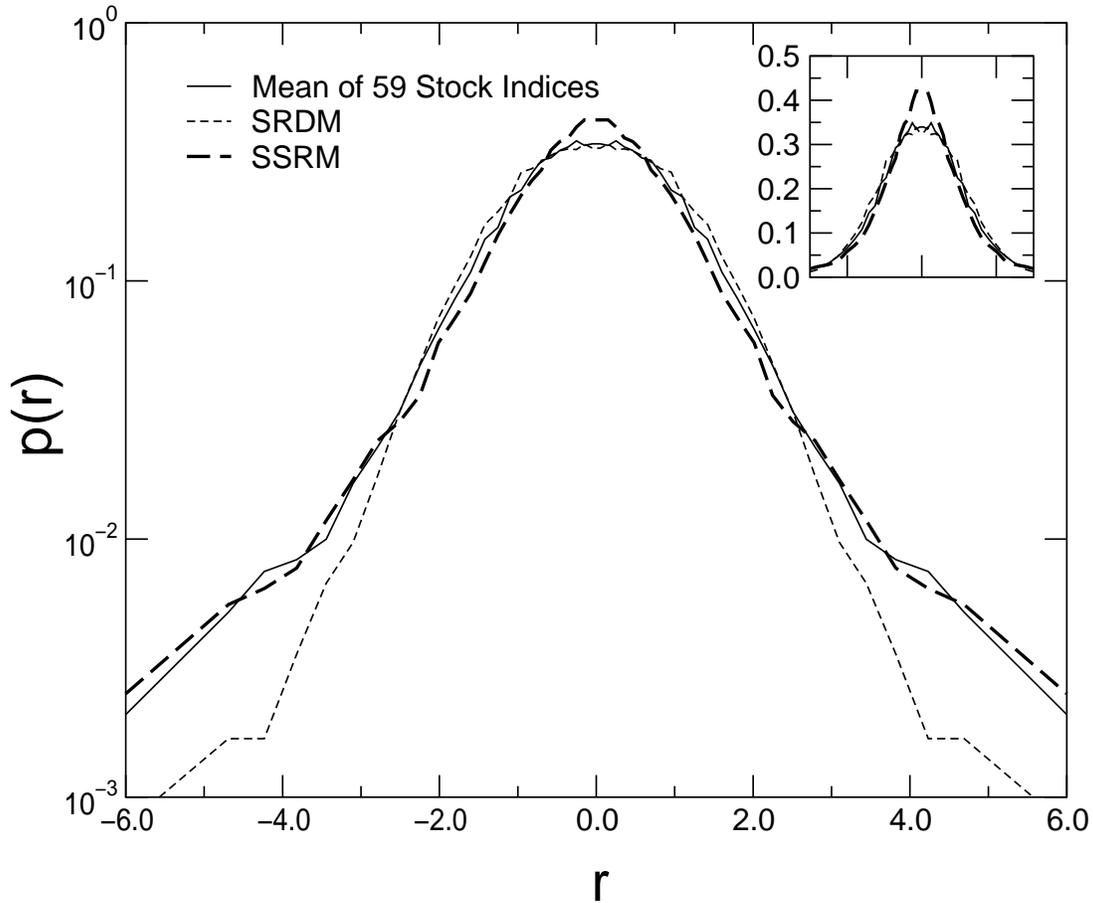}}
\medskip
\medskip
\caption{Symmetrized probability density distribution $p(r)$
of the returns $r_{\tau}$ measured over periods ~$\tau=10$~d for 59
stock indices.  Also shown are the SSRM and SDRM for $\tau=10$~d.
These distributions are typical of $\tau>1$~d.  We find there is a fat
tail in SSRM but not in SDRM, indicating that the origin of the fat
tails lies in known long-range correlations in the absolute returns.
Inset follows a linear (not semilog) scale.  The distribution of $r_1$
has been normalized to unit variance.  For $\tau=1$~d all three
distributions are identical (not shown).  }
\label{hist}
\end{figure}

Even for Gaussian returns, however, the presence of correlations can
lead to anomalous behavior, such that $r_1$ and $r_\tau$ may have
non-identical probability distributions.  We therefore develop a
method to ``subtract'' the effects of correlations. For each of the 59
time series, we generate a modified control time series by shuffling
the order of the daily returns (Fig.~\ref{sp}). This shuffled daily
returns model (SDRM) will have a probability distribution identical to
the real data for $\tau=1$~d, but lacks all correlations. Hence, for
$\tau>1$~d the real data and the SDRM will in general not have
identical distributions (Fig.~\ref{hist}) unless correlations are
lacking.  Thus, we now have a way to test the hypothesis that the fat
tails in $p(r_\tau)$ persist solely due to correlations.  If the
probability density distribution p($r_\tau$) is fat tailed for the
real data but not for the SDRM, then the conclusion would be that the
fat tails in $p(r_\tau)$ are due to correlations.

Shuffling the data destroys all kinds of correlations---indeed, the
 data become independent numbers.  Specifically, shuffling destroys
 the known long-range power law correlations in the volatility of the
 returns as well as the short-range correlations in the signs of the
 returns.  We thus develop a method for ``subtracting'' only the
 correlations in the signs of the returns $r_1(t)$ while preserving
 (volatility) correlations in the absolute returns $|r_1(t)|$
 (Fig.~\ref{sp}).  For each of the 59 time series, we generate a
 second modified control time series by shuffling the order of the
 signs---but not of the absolute values---of the daily returns. This
 shuffled signs return model (SSRM) will have a symmetrized
 probability distribution identical to the real data and to the SDRM
 for $\tau=1$~d, but not necessarily for $\tau>1$~d.

We study the symmetrized probability density distribution function
$p(r_\tau)$ of the returns $r_\tau$ from 59 stock markets and compare
them to those of the SDRM and the SSRM. We focus on the fat tailed
regions of the distributions by studying a properly defined
modified characteristic function
\begin{equation}
\begin{split}
f(\tau)= 
\int dr_\tau~ p(r_\tau)~ \exp{[-(|r_\tau|-r_0)^2]} \\
\simeq
(1/N) \sum_t \exp{[-(|r_\tau|-r_0)^2]}
\;\;,
\end{split}
\end{equation}
where $t$ is time in days.  In order to study the fat tailed region
while retaining good statistics, we chose a value $r_0=5$
corresponding to 5 standard deviations. (We also studied higher
moments, but these are extremely sensitive to large events, rendering
the results not statistically significant.)  Similarly, to study the
central bell curve region, we define a second function
\begin{equation}
g(\tau)= \int dr_\tau~ p(r_\tau) \exp{(-r_\tau^2)} \simeq (1/N)\sum_t
\exp{(-r_\tau^2)}\;\;\; ,
\end{equation}
where $N=\sum_t 1.$ 
In practice, we calculate these functions directly
from the returns, rather than through the distributions, to obtain
better statistics.

We find that the fat tails almost disappear for $\tau> 1$~d in the
SDRM, showing that correlations are necessary for maintaining the fat
tails for $\tau> 1.$ This finding is consistent with the results
reported in ref.~\cite{gopi} and rules out the possibility that the
daily L\'evy-like distribution is responsible for the persistence of
fat tails.  In fact, if this were so, the fat tails would persist for
$\tau>1$~d even after shuffling the order of the returns $r_1(t)$,
contrary to our findings.  One must conclude that the fat tails are
mainly due to correlations.  Note, however, that for the SDRM, the fat
tails do not disappear entirely and $p(r_\tau)$ never becomes truly
Gaussian even for $\tau\rightarrow $100~d (business, not calendar,
days), so a truncated L\'evy distribution of $r_1(t)$ is in principle
not ruled out for $\tau=1$~d~\cite{mantegna,volatility-stanley}.  Also
not ruled out is the distribution suggested in ref.~\cite{lrc1}

Our most important finding is that the fat tails remain 
intact for the SSRM, showing that the fat tails can persist for
$\tau>1$~d when the data lack sign correlations but have long-range
correlated absolute values.  This finding proves that whatever the
choice of the distribution $p(r_1)$ of daily returns, long-range
correlations in the volatility are necessary to explain the behavior
of $p(r_\tau).$

An important consequence of these findings is that great care must be
taken when trying to study the distributions $p(r_\tau)$ independently
of the correlations.  Our findings show that this is true for
$\tau>1$~d, and it is possible that a study of higher frequency data,
with many daily data points, would show similar behavior for
$\tau<$1~d.  The lower renormalization cutoff could conceivably be as
small as the resolution of the data set, even as small as 10~s for a
high volume American stock.

We also find that the central bell curve region of the distribution of
returns is more similar to that of the SDRM than to the SSRM for
$\tau>1,$ showing that in this region the real data are more similar
to a Gaussian and that Markovian sign correlations in the returns are
important in maintaining the Gaussian-like appearance.  Finally, we
also find that the behavior of the distribution $p(r_\tau)$ is
remarkably similar for different $\tau.$

The new results reported here are of broad interest and scientifically
important because long-range correlations and fat tailed distributions
can be found in many physical, chemical, and biological phenomena.
Moreover, the prices of many financial derivative products depend only
on the distribution of returns.  The existence of long-range
volatility correlations and fat tailed distributions underlying
financial time series has been known for some time.  What was not
fully understood is the origin of the fat tails---which turns out to
persist for large lags mainly because of long-range correlations in
the volatility.  Note that exponentially decaying (i.e., not power
law) correlations cannot lead on their own to fat tails at large lags.
A systematic study of the S\&P~500 index by Gopikrishnan {\it et
al.}~\cite{gopi} had found that the observed scaling of the
distributions is due to time dependencies. Here, we have shown that
shuffling only the signs, but not the absolute values, of the returns
allows the fat tails to persist---hence the fat tails are due to
volatility correlations rather than any other kind of time dependency.

\begin{figure}
\centerline{\includegraphics[width=5cm,angle=-90]{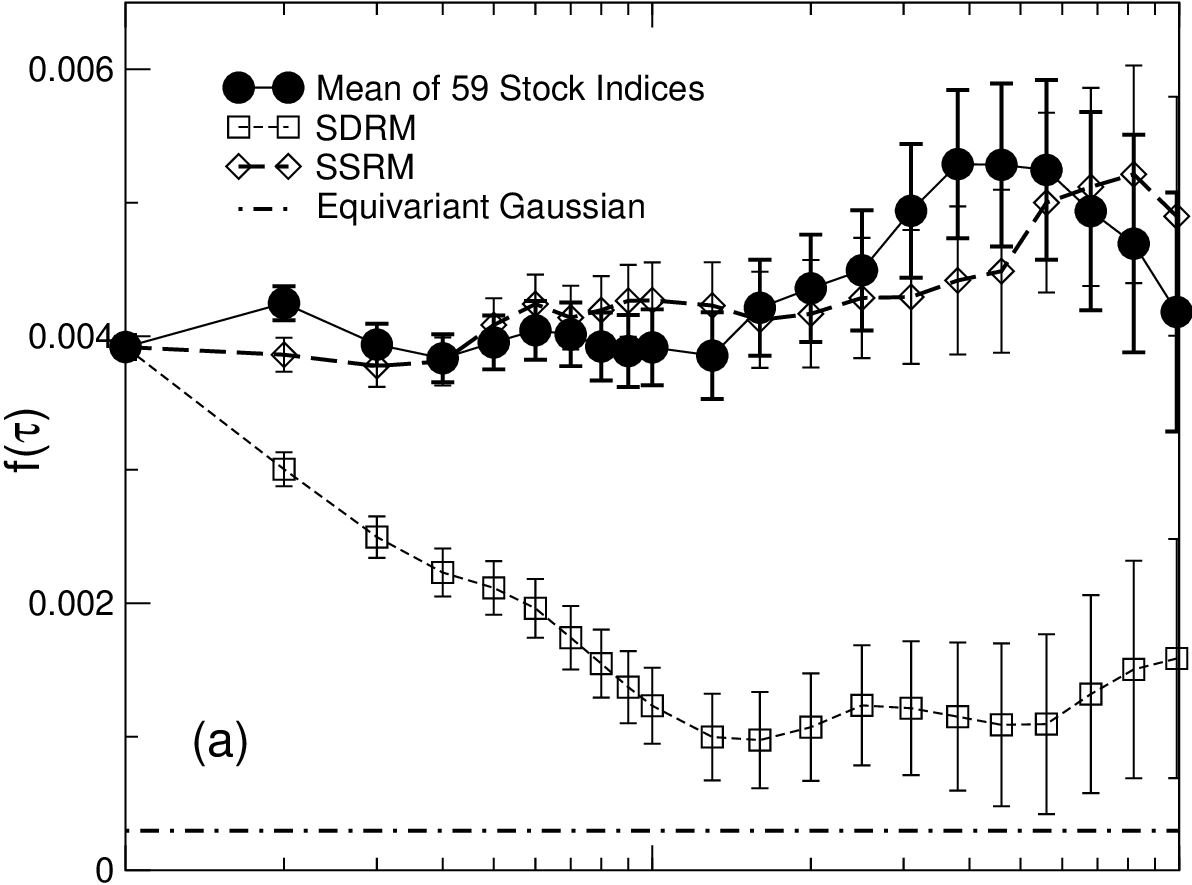}}

\bigskip
\centerline{~~\includegraphics[width=5.6cm,angle=-90]{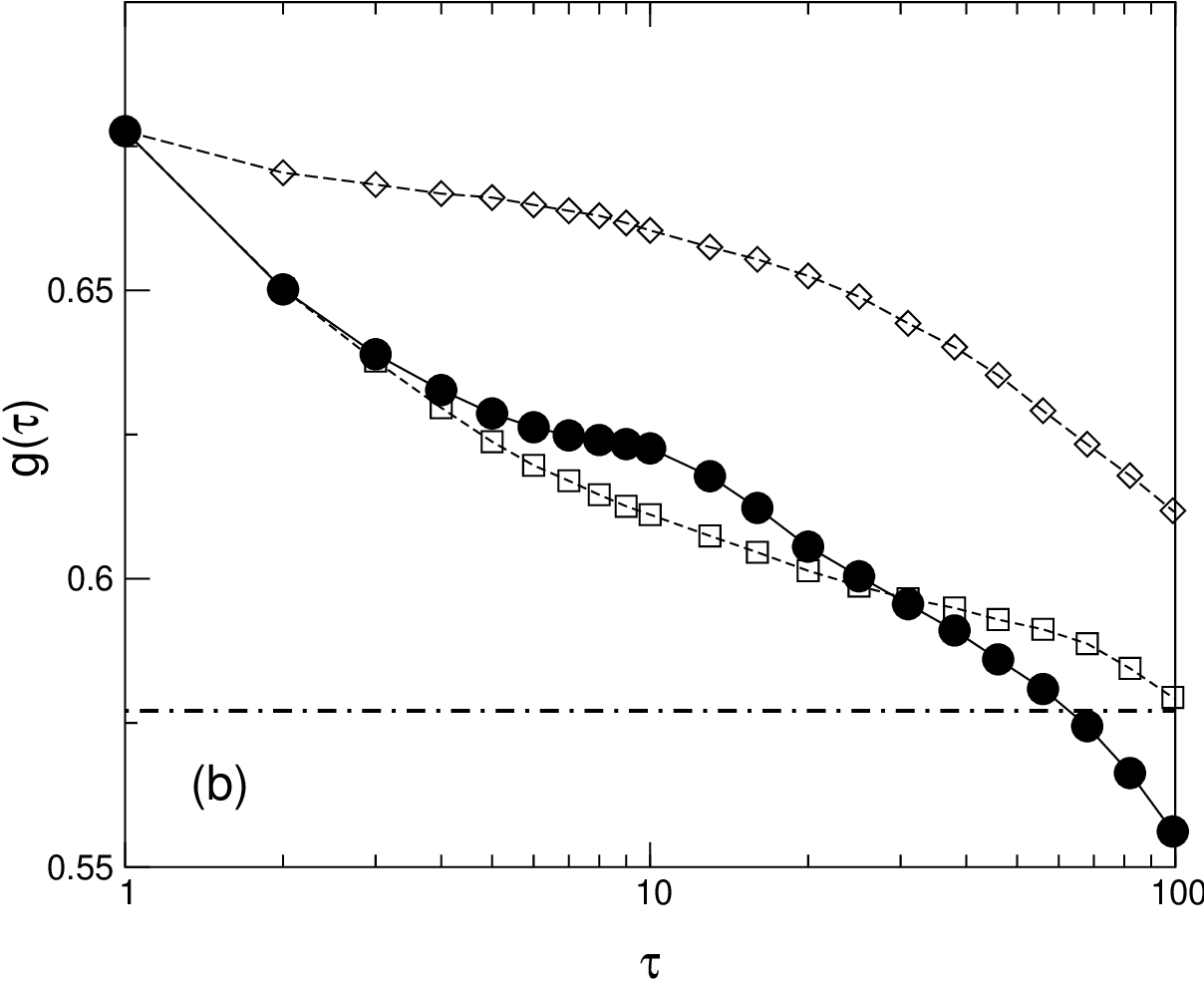}}

\caption{(a)~Mean characteristic function $f(\tau)$ for 59 stock
indices, along with the SSRM and SDRM controls, focusing on the fat
tails.  The values for the SDRM become significantly lower with
$\tau,$ indicating that the tails are less fat for the shuffled data.
The values for the SSRM, however, are remarkably consistent with the
original data, showing that known long-range volatility correlations
are the real cause of the observed non-Gaussian fat tailed
distributions. The loss of the fat tail for the SDRM thus rules out a
true L\'evy stable distribution.  (b)~Mean characteristic function
$g(\tau)$ for the same datasets. Note that the SSRM has many more
returns near zero for $\tau>1$~d, leading to a higher value of $g.$
This result shows that sign correlations in the real data play an
important role that counteract the volatility correlations.  Another
result seen in (a) and (b) is that an equivariant Gaussian
approximation is extremely good for small $r$ (as seen from $\Delta
g/g\simeq 20\%$), but very bad for large $r$ (since $\Delta f/f\simeq
1000\%)$, a finding potentially important for options pricing theory.}
\label{cf}
\end{figure}

We comment on the finding that the real data are more similar to the
SDRM in the central bell curve region, but more similar to the SSRM in
the fat tailed region. Fig.~\ref{cf} shows that the absence of
volatility correlations causes a collapse of the fat tails in the
SDRM, but that the absence of sign correlations causes large
deviations in the central bell curve region in the SSRM, for $\tau>1.$
The real data appear qualitatively somewhere in between the SDRM and
the SSRM.  The implication is that the error of neglecting sign
correlations can somehow ``compensate'' the error of ignoring
long-range volatility correlations for small price changes.  As a
result, for $\tau>1$~d the central region is deceptively well
described by a Gaussian.  The two errors cancel each other out, hence
the success of Bachelier's century-old Gaussian theory of stock
returns.  The Gaussian assumption, known to be only approximate, was
nevertheless the basis of the Economics Nobel Prize in 1997 (awarded
for the work that led to the Black-Scholes Options Pricing Theory).
The main flaw in the Gaussian theory is that it
cannot explain the fat tails that point to relatively rare but large
events, such as the great correction of 1987.  There are other known
phenomena in which the cancellation of two errors has led to
surprisingly good models, a classic example being the original theory
of polymer melts by Flory~\cite{flory}, in which the error in
estimating repulsive and attractive energies could effectively
``cancel'' each other out, such that the model became better than
otherwise would be expected.

In summary, our findings indicate that the fat tailed distributions of
stock returns are mainly due to volatility correlations.  More
generally, we have shown that fat tailed distributions can arise from
long-range correlations in the absolute value of any time series.  We
note that the shuffling techniques we have developed here are general
and can be applied to the study of time series generated by other
dynamically rich complex systems that present similar challenges.

We thank the Brazilian agencies CNPq (for support) and CAPES (for
funding the visit to Brazil of MS) and P. Gopikrishnan, P. Ch. Ivanov
and H. Eugene Stanley for very helpful comments.

\begin{thebibliography}{mt1}

\bibitem{econophysicsbook} R. N. Mantegna and H. E.  Stanley, {\it An
Introduction to Econophysics: Correlations and Complexity in Finance}
(Cambridge University Press, Cambridge, 2000).

\bibitem{montroll}
%
E. W. Montroll and W. W. Badger, {\it Introduction to Quantitative
Aspects of Social Phenomena} (Gordon and Breach, New York, 1974).



\bibitem{shlesinger} M. F. Shlesinger, G. Zaslavsky and U. Frisch,
eds., {\it L\'evy Flights and Related Topics in Physics} (Springer,
Berlin, 1995).


\bibitem{prl1} J.-P. Bouchaud, A.  Matacz and M.  Potters,
Phys. Rev. Lett. {\bf 87}, 228701 (2001).

\bibitem{prl2} V. M.  Eguiluz and  M.  G.  Zimmermann,
Phys. Rev. Lett. {\bf 85}, 5659 (2000).


\bibitem{plamen1} B. Podobnik, P. Ch. Ivanov, Y. Lee, A. Chessa and
H. E. Stanley,
%
%
Europhys. Lett. {\bf 50(6)}, 711
%
(2000).

\bibitem{plamen2}  
B. Podobnik, P. Ch. Ivanov, Y. Lee and H. E. Stanley,
%
%
Europhys. Lett. {\bf 52}, 491
%
(2000).



\bibitem{lrc1}
M. Pasquini and M. Serva,
%
%
Euro. Phys. J. B {\bf 16}, 195 (2000).

\bibitem{lrc-old0} Y. Liu, P. Gopikrishnan, P. Cizeau, M. Meyer,
C.-K. Peng and H. E. Stanley, Phys. Rev. E. {\bf 60}, 1390 (1999).

\bibitem{gopi} 
P. Gopikrishnan, V. Plerou, L. A. N. Amaral, M. Meyer,
and H. E. Stanley, 
%
%
Phys. Rev. E {\bf 60,} 5305
%
 (1999).


\bibitem
{lrc2}
M. Pasquini and M. Serva,
%
%
%
Econ. Lett., {\bf 65}, 275 (1999).



\bibitem
{mantegna2}
%
R. N. Mantegna and H. E. Stanley, {Nature} {\bf 383}, 587 (1996).


\bibitem
{mantegna}
%
R. N. Mantegna and H. E. Stanley, {Nature} {\bf 376}, 46 (1995).




\bibitem {mandelbrot} B. B. Mandelbrot, J. Bus. {\bf 36}, 394
(1963).

\bibitem{kadanoff} 
%
L. P. Kadanoff, {Simulation} {\bf 16}, 261 (1971).


\bibitem
{mantegna3}
%
R. N. Mantegna, { Physica A} {\bf 179}, 232 (1991).

\bibitem{Li} 
%
W. Li, { Int. J. Bifurcation and Chaos} {\bf 1}, 583 (1991).



\bibitem
{takayasu}
%
H. Takayasu, H. Miura, T. Hirabayashi and K. Hamada, {Physica A} {\bf
184}, 127 (1992).




\bibitem
{bak}
%
P. Bak, K. Chen, J. A. Scheinkman and M Woodford, {Ricerche
Economiche} {\bf 47}, 3 (1993).



\bibitem
{bouchaud}
%
J.-P. Bouchaud and D. Sornette, {J. Phys. I France} {\bf 4}, 863
(1994).




\bibitem
{ghashghaie}
%
S. Ghashghaie, W. Breymann, J. Peinke, P. Talkner and Y. Dodge,
{Nature} {\bf 381}, 767 (1996).






\bibitem
{levy} 
%
M. Mevy and S. Solomon, {Int. J. Mod. Phys. C} {\bf
7}, 595 (1996).


\bibitem
{galluccio}
%
S. Gallucio, G. Caldarelli, M. Marsili and Y.-C. Zhang, {Physica A}
{\bf 245}, 423 (1997).



\bibitem
{bak2}
%
P. Bak, M. Paczuski and M. Shubik, {Physica A} {\bf 246}, 430 (1997).



\bibitem
{vandewall}
%
N. Vandewalle and M. Ausloos, {Physica A} {\bf 246}, 454 (1997).


\bibitem
{Liu}
%
Y. Liu, P. Cizeau, M. Meyer, C.-K. Peng and H. E. Stanley, {Physica
A} {\bf 245,} 437 (1997).



\bibitem
{cizeau}
%
P. Cizeau, Y. Liu, M. Meyer, C.-K. Peng and H. E. Stanley, {Physica A}
{\bf 245}, 441 (1997).

\bibitem{volatility-stanley}

B. Podobnik, K. Matia, A. Chessa, P. Ch. Ivanov, 
Y. Lee and  H. E. Stanley, Physica A {\bf 300}, 300 (2001).


\bibitem
{ingersoll} 
%
J. E. Ingersoll, {\it Theory of Financial
Decision Making \/} (Rowman \& Littlefield, Savage, 1987).


\bibitem
{bacchelier} 
%
L. Bachelier, 
%
Ann. Sci. \'Ecole Norm. Sup. {\bf 17}, 21 (1900).


\bibitem{tlf} 
R. N. Mantegna and H. E. Stanley, Phys. Rev. Lett. {\bf73}, 2946 (1994).


\bibitem{lrc-old1}
M. Greene and  B. Fielitz, J. Financial Econ. {\bf 4}, 339 (1977).

\bibitem{lrc-old2}
V. Akgiray, J. Bus. {\bf 62}, 55 (1989).



\bibitem{lrc-old3} A. W. Lo, Econometrica {\bf 59},1279 (1991).

\bibitem{lrc-old4}
N. Crato and  P. Rothman, Econ. Lett.   {\bf 45}, 287 (1994).







% \bibitem{indices} 
% Obtained from yahoo.com:
% AEX, AORD, ATG, ATX, BFX, BSESN, BVL30, BVSP, CCSI,
% DJA, DJI, DJT, DJU, DOT, FCHI, FTSE, HEX, HSI, IBC, IGRA, IIX, IPSA,
% IXIC, JKSE, KFX, KLSE, KS11, KSE, MERV, MID, MTMS, MXX, N225, NDX,
% NTOT, NYA, NZ40, OEX, PSE, PSI, PX50, RUA, RUI, RUT, SAX, SETI, SML,
% SMSI, SOOX, SPC, SSEC, SSMI, STI, TA100, TSE, TWII, VLIC, XMI, XU100.




% \bibitem{fn1} We also studied higher moments, but these are extremely
% sensitive to large events, rendering the results not statistically
% significant.

% \bibitem{nobel} The prize was awarded for the work that led to the
% Black-Scholes Options Pricing Theory.


\bibitem
{flory}
%
P. J. Flory, {\it Principles of Polymer Chemistry\/} (Cornell
University Press, New York, 1953).







\end{thebibliography}
\end{document}